# APPROXIMATING NASH EQUILIBRIUM UNIQUENESS OF POWER CONTROL IN PRACTICAL WSNS

Evangelos D. Spyrou and Dimitrios K. Mitrakos

School of Electrical and Computer Engineering, Aristotle University of Thessaloniki. Thessaloniki, Greece

## *ABSTRACT*

*Transmission power has a major impact on link and communication reliability and network lifetime in Wireless Sensor Networks. We study power control in a multi-hop Wireless Sensor Network where nodes' communication interfere with each other. Our objective is to determine each node's transmission power level that will reduce the communication interference and keep energy consumption to a minimum. We propose a potential game approach to obtain the unique equilibrium of the network transmission power allocation. The unique equilibrium is located in a continuous domain. However, radio transceivers accept only discrete values for transmission power level setting. We study the viability and performance of mapping the continuous solution from the potential game to the discrete domain required by the radio. We demonstrate the success of our approach through TOSSIM simulation when nodes use the Collection Tree Protocol for routing the data. Also, we show results of our method from the Indriya testbed. We compare it with the case where the motes use Collection Tree Protocol with the maximum transmission power.*

## *KEYWORDS*

*Transmission Power, Packet Reception Ratio (PRR), Game Theory, Distributed Optimisation, Potential Game*

## 1.INTRODUCTION

Wireless sensor networks are widely deployed in a variety of environments, supporting military surveillance [1], emergency response [2], and scientific [3], due to their sensing and communication capabilities. The environmental diversity in conjunction with resource constraints of wireless sensor devices makes reliable and energy-efficient wireless communication a difficult issue.

Power control algorithms for Wireless Sensor Networks (WSN)s are essential, in order to reduce energy consumption. This is appropriate for a plethora of applications [4]. The problem of adjusting nodes' transmission power levels is a major issue in resource limited devices due to low signal bandwidths, uncertainty or non-linearity of system models. In order to lengthen the lifetime of the network, WSN algorithms take advantage of platform-based design techniques to increase energy-efficiency in a coordinated manner [5]. On the other hand, reducing transmission power may have a major impact on the transmission reliability of nodes.

The use of maximum transmission power improves reliability [6,7] ; however, this results to high energy consumption. WSN radio transceivers, such as CC2420 [8] offers the TXCTRL.PA_LEVEL register to specify the transmission power level during runtime. The datasheet provides the designer with a discrete set of 8 transmission power levels in the range of -25 dB. to 0 dB, in order to optimise an objective function, such as reliability will be maximised and energy consumption will be kept to a minimum level.

Intuitively, we can consider each node as a selfish player that wants to maximise its profit with a low cost in a network of other players. Furthermore, nodes exchange their actions, in order to provide feedback on the loss or profit defined by a global function. In this manner, nodes can dynamically react to changing conditions and energy budget, in order to optimise their utility. Hence, we can use game theoretic concepts to model the behaviour or rationality of the nodes in a distributed manner. Game theory has been used in a plethora of optimisation problems, including power control, resource allocation and bandwidth optimisation in [9] and references therein. The heart of this research is to determine the unique and optimal transmission power levels of the nodes with which energy consumption is minimised and transmission reliability is maximised. In theoretical approaches, optimisation is undertaken in the continuous domain, where certain assumptions provide optimality and uniqueness of the fixed point [10]. However, in practical scenarios continuous optimisation is not the most appropriate solution, since the optimal solution may reside in between two discrete values. Moreover, wireless sensor devices avoid the support of floating point transmission power levels.

In this paper, we propose a game theoretic model that optimises the trade-off between energy consumption, and packet delivery performance. We present our model and prove that this game is a potential game [11]. Potential games are games where the incentive of players to change their strategy can be expressed in one global function, the potential function. This proves that there is an equilibrium point. We give the sufficient conditions to show that the equilibrium point is unique. We construct an algorithm and interface it with the Collection Tree Protocol (CTP) [12] and evaluate its performance and energy-efficiency. We show simulation results of the performance and energy consumption of our solution. Thereafter, we compare it with the discretised version of our algorithm that approximates the optimal solution to the nearest discrete value. In order to do that, we assume that the TXCTRL.PA_LEVEL register values and the signal amplitude are linearly related, and we discretise the transmission power range to the granularity of 1 dB. This gives us a larger number of transmission power levels [13].

The contributions of this paper are the following:

- We determine a unique equilibrium solution for our proposed power control game. We provide simulation results that show the convergence of our algorithm in a finite set of iterations.
- We discretise the algorithm and provide a comparison via TOSSIM [47] simulations and show the difference in the PRR for both versions of the algorithm. We also compare our result with the use of maximum transmission power, where performance is maximised and we show that our solution approaches it while reducing the network's energy consumption.
- We put the discretised algorithm on all nodes on Indriya [14] and show the difference in link performance with an algorithm running at maximum transmission power. The results show that the difference is approximately 10% while the energy is minimised.
- We show that our algorithm even, obviously more energy efficient, it does not lack a great deal in link quality, since it includes a large number of good links.

The paper includes the following sections: Section 2 presents the related work, Section 3 gives a brief introduction to game theory, Section 4 formally describes the power control algorithm, Section 5 provides the necessary information regarding the utility function we define. Furthermore in Section 5, we give substantial evidence regarding the convergence and optimality of the solution, Sections 6 and 7 shows the experimental results obtained and Section 8 presents the conclusions and future work.

## 2. RELATED WORK

There has been significant research on power control for wireless networks over the past years [28, 29, 30, 31, 32 33]. However, most recent topology control algorithms deal with adapting transmission power, in order to meet specific targets, such as high PRR and low energy efficiency. Some of them are presented after in this section. We have to highlight that we mainly focused on the practical related works and others that use game-theoretic solutions. Furthermore, our approach uses a node-based neighbourhood link quality optimisation in comparison to other papers.

Nahir et al. [34], provided a game-theoretical solution to the topology control problem, by addressing three major issues: the price of establishing a link, path delay and path congestion proneness. They established that bad performance due to selfish play in the considered games is significant, while all but one are guaranteed to have a Nash equilibrium point. Furthermore, they showed that the price of stability is typically 1; hence, often optimal network performance can be accomplished by being able to impose an initial configuration on the nodes. Note that this work is based on directed graphs, where bidirectional links do not exist, which makes the solution not unsuitable for practical WSN deployments. Furthermore, the authors express their concern regarding the computational tractability of their solution. Finally, one of the games they consider did not admit a Nash equilibrium and further investigation on this system is essential, in order to examine its behaviour.

Komali et al. [35], analysed the creation of energy efficient topologies with two proposed algorithms. Specifically, their game-theoretic model specified that nodes have the incentive to preserve connectivity with a sufficient number of neighbours and that the network will not partition. They proved that their game is an exact potential game and that a subset of the resulting topologies is energy efficient. They addressed the major issue of fair power allocation by providing the argument of efficient allocation vs fair allocation. This work is mathematically solid; however, it did not take into account link quality or network performance in their optimisation problem. Finally, they provided simulation results only to evaluate their work.
The characteristics and behaviours of wireless links are now more understood. There has been work measuring the effects of varying power levels and showing the irregularity of radio ranges and the lack of link symmetry [36, 37]. The relationship between PRR and RSSI for the Chipcon CC2420 radio was established in [38]. Subsequent work then looked at the differences in behaviour between indoor and outdoor networks, and fluctuations in link quality over longer durations of time [39].

Regarding of Topology Control specifically, [39] contributes a comprehensive review of this field which we summarise. Given the diversity of link behaviours influenced by their environment, experimentation for much of the early TC work was carried out using graph theory and simulation studies for tractability reasons. Yet, this work did not consider aspects like realistic radio ranges, node distributions or node capability/capacities into account, limiting their usefulness for real sensor networks [40, 41, 42]. For example, some have assumed that link costs are proportional to link length, but in reality a more complex relationship is evident [36, 37].

The main competitors in the practical Topology control area are PCBL [36] and ART [39, which we introduce next. PCBL was derived from link quality observations showing that links with a very high PRR remain quite stable. They then categorise links as blacklisted, middling or highly reliable. The power in the latter is minimised to their lowest stable power setting while the blacklisted are not used at all. The middling links are those that lie between the two and are set to full power. Given the expense of probing the network to establish the link categories, this protocol cannot work with dynamic routing protocols such as CTP. CTP aims to find the least expensive routes through the network. To overcome such link probing, link quality metrics have been used to approximate PRR in ATPC. Specifically there is a link between RSSI and PRR, and LQI and PRR over a monotonically-increasing curve. Further, linear correlations between transmission power levels and RSSI/LQI are observed at the receiver but are different for each environment monitored. Therefore ATPC estimates the slope and uses closed feedback to adjust the model to the current situation to achieve lower bound RSSI (PRR). However, RSSI is not a good enough link quality indicator, since it is highly affected by obstructions in between nodes and radio propagation issues.

Hackmann et al., showed that RSSI and LQI cannot always realistically estimate PRR in indoor environments, nor can instantaneous probing represent the behaviours of a link over time. They do not propose ART, which does not rely on estimates of link quality nor does it involve long bootstrapping phases. Being more dynamic, ART adapts link power to changes in the environment as well as contention using a gradient. Also, where applications expect acknowledgment messages, ART can piggyback these to reduce communication overhead. ART selects the appropriate transmission power based on the failures observed when the target PRR is 95% and a contention gradient. However, a given node in ART will have to switch between power settings to communicate to its neighbourhood nodes. This has potential issues in terms of scaling.

## 3. GAME THEORY AND POTENTIAL GAMES

Game theory studies mathematical models of conflict and cooperation [15] between nodes, in our case. The rationality of a node is satisfied if it pursuits the satisfaction of its preferences through the selection of the appropriate strategies. According to decision theory, the preferences of the decision maker (node) need to satisfy some general rationality axioms. Then, its behaviour is described by a utility function. Utility functions provide a quantitative description of the node's preferences in different stages of the game. The main objective of each decision maker is the maximization of his/her utility function, which results in the identification of a Nash equilibrium [16]. The meaning of the Nash equilibrium is the fact that after the exchange of strategies between players, none of them has the benefit of deviating, in order to increase its payoff. Games may contain one or more Nash equilibrium points, or none at all.

In 2008, Daskalakis et al., [17] proved that finding a Nash equilibrium is PPAD-complete. Polynomial Parity Arguments on Directed graphs (PPAD) is a class of total search problems for which solutions have been proven to exist; however, finding a specific solution is quite difficult if not intractable. Note that PPAD-complete problems provide weaker evidence of intractability than NP-complete, even though a solution is unlikely to be found.

This development altered the perception of researchers aiming to employ game theory in their individual problems, thereby leading them to a specific class of games called 'Potential Games', due to their important properties, which are that pure equilibria always exist and best response dynamics are (almost surely) guaranteed to converge. Potential games are classified as exact and ordinal potential games.

We are interested in the exact potential games and refer the reader to Monderer's paper for details on the ordinal class. Given the same specification as in the exact potential game, V(A) has the same behaviour with the exact potential function. More specifically, the function tracks the actual improvement.

## 4. POWER CONTROL ALGORITHM

In this section we provide analysis on the non-cooperative game that we propose. We prove that the game is a potential game and give the necessary conditions for the existence of a unique Nash Equilibrium.

### 4.1. SYSTEM MODEL

We consider a set of nodes M = {1,....,i} and denote by their power allocations by P = ($p_1$,...$p_M$). We assume that the nodes are randomly distributed, whereby the density of the network is homogeneous across the WSN deployment area. Also, we assume that each node i can adjust its transmission power $p_i$ within a range and the P is a convex and compact set.

$$0 \leq p_i^{min} \leq p_i^{max} \qquad (1)$$

Even though we are aware that in practice a mote has a discrete set of transmission power levels, with which it can transmit, we assume that any power within the above range can be sustained.

### 4.2. CONNECTIVITY

We assume that the density of the network is homogeneous across the deployment area. Since many WSNs are randomly deployed as in Indriya, we can consider a WSN as a random graph. In [18] and references therein, a plethora of protocols are presented that measure connectivity probabilistically based on the necessary node degree. Specifically, according to established random graph theory, [19, 20], if the number of neighbours (degree) of every node exceeds the threshold 5.1774log(N) [21], the network is asymptotically connected as N increases, meaning that there is a path to the sink from any node in the network. Furthermore, in [22], the authors propose a protocol that requires 9 neighbours to satisfy connectivity with a probability of 0.95 for a resulting symmetric graph, when the network size is from 50-500 nodes.

In this paper we consider the small-world Model A from [23], where there are M nodes in the network and each one arbitrarily selects m nearest neighbours to connect to. Essentially, we utilise the variant of this small-world model, where node locations are being modelled by a stochastic point process. The number m of neighbours consists of nearest neighbours and shortcuts. A shortcut is an edge between two nodes if either of the two nodes exists in the nearest neighbour set of the other. If a node is connected by a nearest neighbours and a shortcut, multiple edges are replaced by a single one. The presence of the shortcuts reduces the network diameter. Furthermore, we have to note that m is the number of neighbours a node has in terms of a spatial graph, and (N-1)p is the number of neighbours it has via shortcuts. The authors dictate that to ensure connectivity the quantities $m = (1+\delta)\sqrt{2\backslash log(M)}$ and $Mp = (1+\delta)(\sqrt{2\backslash log(M)}$, where δ > 0, are sufficient. Hence connectivity is preserved with a smaller degree of (nearest neighbours plus shortcuts). We select a degree of 6 for each node. It is well known that the node degree can be reached by adjusting the transmission power; hence, we denote $PT_i$ as the transmission power level, which satisfies the condition more than 6 nodes exist in the neighbourhood of each node. It is intuitive, by the nature of CTP that each node will select shortcuts to forward its data towards the sink, based on its Expected Transmission Count (ETX) values with the forwarding nodes.

## 4.3. NEIGHBOUR COMMUNICATION RELIABILITY (NCR)

For a wireless link (i,j), the Packet Reception Ratio $PRR_{i,j}$ is defined as the ratio of the number of packets received by node j over the number of packets sent by node i. To measure the reliability of links around node i, we define a new metric called Neighbour Communication Reliability ($NCR_i$) as

$$NCR(p_i, p_{-i}) = \frac{\sum_{j \exists N_i(p_i, p_{-i})} PRR_{ij}}{\bigcup_{j \exists N\_i(p_i, p_{-i})} N_j(p_i, p_{-i})} \quad (2)$$

where $p_i$ is the power level of node i, $p_{-i}$ means the power levels of all neighbours N except i, $N_i(p_i, p_{-i})$ is the set of nodes such that $\forall j \exists N_i(p_i, p_{-i}), PRR_{ij} > 0$. In practice, every node i can obtain $NCR_i$ at run time by every node $j \exists N_i$, j calculating the average $PRR_{ij} \exists N_j$ and periodically broadcasting average $PRR_{ij}$, which is derived as follows

For any link $(i,j) \exists N$. $PRR_{ij}$ can be expressed by approximation as:

$$PRR_{ij} = (1 - BER)^{8f} \quad (3)$$

Where f is the packet length in bits. The Bit Error Rate (BER) [24] is

$$BER_{ij} = \frac{1}{2}(1 - \sqrt{\frac{SINR_{ij}}{1 + SINR_{ij}}}) \quad (4)$$

where $SINR_{i,j}$ is the Signal-to-Interference-plus-Noise Ratio (SINR) of the transmission from node i to node j. $SINR_{i,j}$ is given by

$$SINR_{ij} = \frac{H_{ij} p_i}{\sum_{t \neq i, t \neq j} H_{tj} p_j + N_0} \quad (5)$$

where $N_0$ is the white noise and $H_{i,j}$ is the channel gain of the wireless link (i,j). Due to the path loss, the larger the distance between nodes i and j the smaller the $H_{k,j}$. We focus on static WSN, hence, we assume that the channel is slow fading in nature and the channel gain of every link remains constant before the convergence of the power control algorithm.

In Figure 1 we provide the reader with the NCR for a given node within the network.

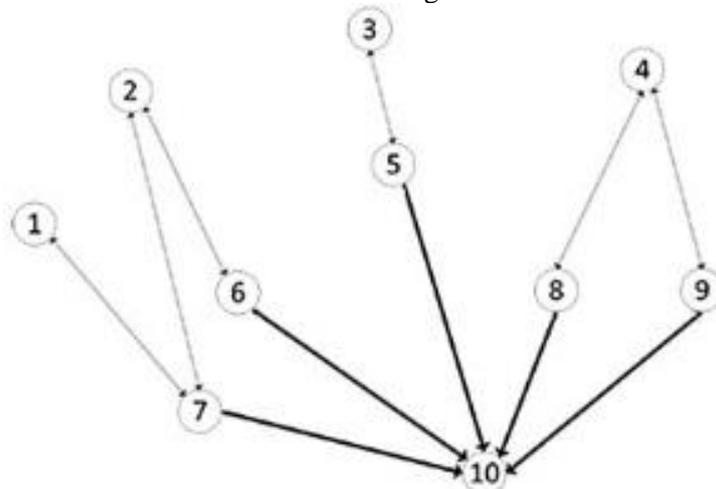

Figure 1. NCR Example

From Figure 1 we see that the NCR is being computed as the incoming PRR obtained from the 1-hop neighbours of node 10, where the links are bold and thicker. Nodes 1-4 do not participate in the NCR calculation.

## 5. UTILITY FUNCTION AND EQUILIBRIUM

Essentially, we need to solve the following optimization problem

$$\max \; \log(1 + 9NCR_i) - \left(\frac{p_i}{25}\right)^2 \quad (6)$$

Subject to
$$PT_i \geq PT_{thr}$$
$$0 < p^{min} < p_i < p^{max}$$

Initially, we formulate our problem as a strategic non-cooperative game

$$\Gamma = \left\{ N, A, \log(1 + 9NCR_i) - \left(\frac{p_i}{25}\right)^2 \right\}, i \ni M \quad (7)$$

Where M is the number of players (nodes in the network), A is the set of strategies given by $A = \{0 < p^{min} < p_i < p^{max}, \forall i \ni M$ and $\log\left(1 + 9NCR_i - \left(\frac{p_i}{25}\right)^2\right)$ is the utility of each player i, which acts as its payoff function.

We denote $\left(\frac{p_i}{25}\right)^2$ as $p_i$, $\log(1 + 9NCR_i)$ as $NCR_i$ and define the utility function of each node i as

$$u_i(p_i) = \begin{cases} NCR_i - p_i & \text{if } PT_i \geq PT_{thr} \\ -p_i & \text{otherwise} \end{cases} \quad (8)$$

We can observe that the utility function is designed to be normalised to 1.

Note that the utility function is strictly concave in $p_i$ and that the feasible strategy space remains A; hence, the existence of the Nash equilibrium is guaranteed [25].

**Theorem 1**: *The game $\Gamma = <N, A, u>$ with the individual utility function (8) is an exactl potential game, where N is the set of all sensor nodes.*

**Proof**: The potential function is given by

$$V(p) = \begin{cases} \sum_m (NCR_i - p_i) & \text{if } PT_i \geq PT_{thr} \\ \sum_m -p_i & \text{otherwise} \end{cases} \quad (9)$$

The fact that $\Gamma$ is a potential game comes as a result by taking the characterisation of the potential games in [11] where $\frac{\partial V(p)}{\partial P_m} = \frac{\partial U_m}{\partial p_m} \ni M$    □

Further, we are able to show uniqueness of the equilibrium point by taking advantage specific properties of the potential function.

**Proposition 1**: The potential game Γ has a unique Nash equilibrium.

**Proof**: Initially, take into account that the potential function is strictly concave and continuously differentiable. It is shown in [26] that in in this case and if the nodes' strategy spaces are convex, which in our case are the intervals, the set of Nash equilibria coincides with the maximizers of the potential function. Since the potential function is strictly concave over a convex strategy space, the maximizer is unique; hence, the Nash equilibrium is unique.

### 5.1. BEST RESPONSE DYNAMICS

In a non-cooperative game, best response dynamics are defined in the context that each player updates its strategy to maximize its utility, given the strategies of other players. In this work, we denote $\xi(i): P_{-m} \to p_m$ as the best response mapping for the i$^{th}$ user, which satisfies

$$\xi(i)(p_{-i}) = argmax_{pi}\, ui(p_i, p_{-i}) \quad (10)$$

where $p_i \exists P_i$. In our potential game, best response dynamics of any player can be acquired by the maximization of the potential function V. Thus,

$$\xi_i(p_{-i}) = argmax_{pi} V_i(p_i, p_{-i}) \quad (11)$$

Since the potential function V is continuous differentiable and strictly concave on P, the strategy space is closed and convex, we can utilise the nonlinear Gauss-Seidel algorithm, which is then guaranteed to converge to the maximum of P [27]. As we have seen from Proposition 1, the game admits a unique NE, thus the algorithm will converge to the unique maximizer of P. The updates of the nodes' strategies running the algorithm are

$$p_i^{\,t+1} = \xi(p_1^{\,t+1}, \ldots, p_{i-1}^{\,t+1}, p_{i+1}^{\,t}, \ldots, p_i^{\,t}) \quad (12)$$

Algorithm (1) shows our algorithm for the game play at a node i.

**Power Control Algorithm at node i**

*BEGIN ALGORITHM*
*REQUIRE pi$^{(0)}$*
*REQUIRE NCR(p$_i$ $^{(0)}$)*
*FOR{j=1 to N$_{iter}$}*
*IF {d > (m+ Np)}*
    *\STATE {u$_i$ = NCR$_i$ - p$_i$}*
 *ELSE*
  *\STATE {u$_i$ = -p$_i$}*
 *ENDIF*
  *ENDFOR*
*\STATE {ξ$_i$(p$_{-i}$) = argmax$_{pi}$ V$_i$(p$_i$,p$_{-i}$)}*
*END ALGORITHM*

## 6. SIMULATION RESULTS

We setup an experiment on TOSSIM, in order to show the difference in performance and energy-efficiency between the continuous and the discretised versions of our algorithm. Initially, we had to add the transmission power change functionality to TOSSIM, which is not provided. Thus, we added two simulation functions in order to ship the transmission power used by the game (in dBs) to the TOSSIM function that puts the packet on the air. In order to do that, one of the functions set the transmission power (in dBs) for any given node running the algorithm. The second function obtained the transmission power used by transmitting node and the transmission power of the destination node, in order to calculate the gains.

We generated a topology from TOSSIM LinkLayerModel tool with 80 nodes, randomly distributed and we set the path loss exponent [45] to 3.3, in order to simulate an indoor scenario. We setup our algorithm to work with CTP, hence we tweaked the Link Estimator to provide each node's PRR every 5 packets. We simulated the continuous and discretised versions of our algorithm and compared them with CTP transmitting without power control, where each nodes were set to transmit at maximum power.

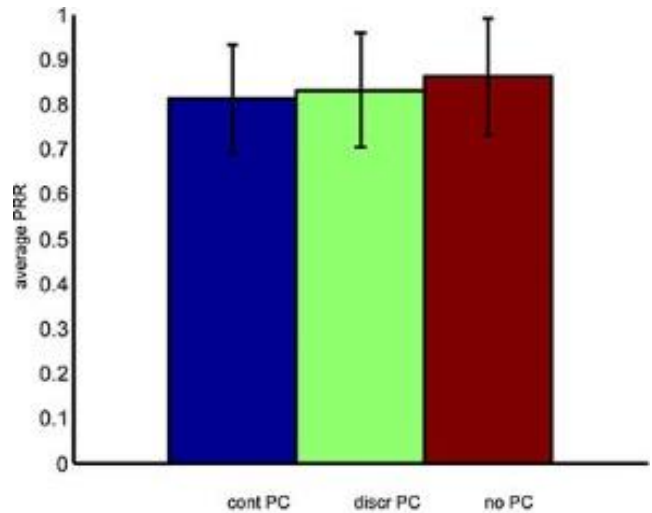

Figure 2. Average PRR

In order to make the continuous space into discrete, we created a simple linear mapping function between the transmission powers in dBs and the IDs accepted by tinyos for the CC2420. we partitioned the transmission power range at the granularity of 1 db. This essentially, is similar to employing a uniform scalar quantiser [46]. The simulation environment allowed us to provide 25 values representing the range of -25 to 0 dB.

International Journal of Computer Networks & Communications (IJCNC) Vol.7, No.6, November 2015

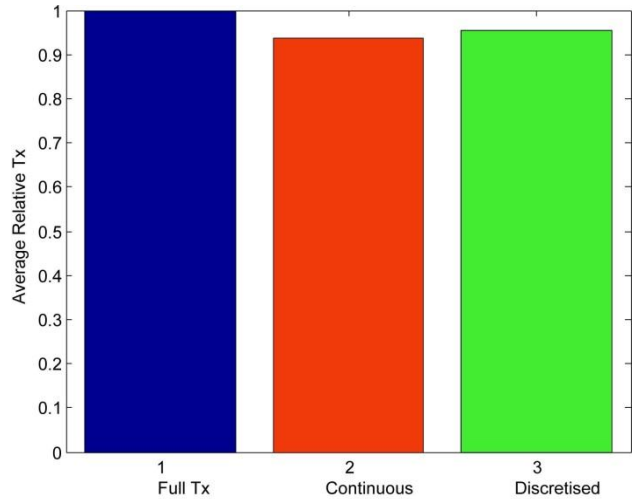

Figure 3. Average Relative Energy

Figure 2 shows the average PRR accomplished by continuous, discretised running the game theoretic algorithm and full Tx. We can determine that the difference is not that significant. In the continuous case the average PRR is 81%, whereas in the discretised version of the algorithm PRR increases by 3%. When no power control is utilised and nodes are transmitting at full power the packet delivery performance is 86%, which is expected since the SINR will be high. However, we also need to investigate the energy consumed for achieving such PRR values.

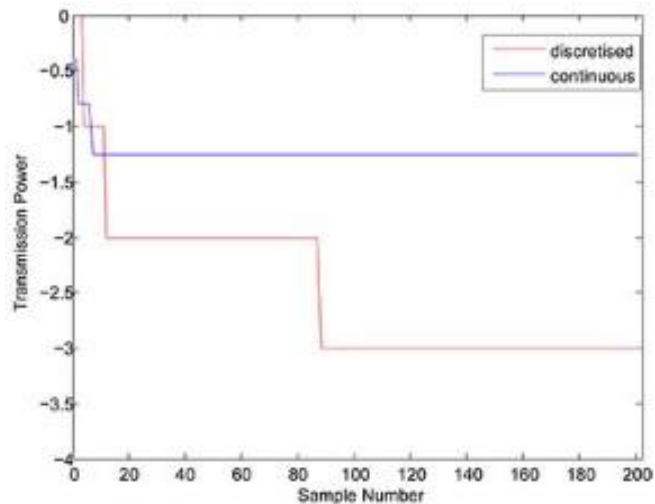

Figure 4. Discretised and continuous Tx values of node 30

We can see from Figure 3 that the average relative energy consumed in all three cases is quite similar. We derived average relative Tx values for all three algorithms by counting the messages sent with each transmission power and we normalised them to 1. Specifically, there is a small difference of less than 10 % between full power and the continuous and discretised algorithms. This shows that nodes running CTP without power control converge to high transmission powers. This may be the case due to the retransmissions of CTP that enhance performance even in the case of collisions. Specifically, we used the default number of retries. Note that CTP provides 30 message retries by default, given in the CTP code. Hence, the algorithm converges to high transmission powers, due to the high transmission reliability.

As an example, we obtained the Tx values of node 30 throughout the experiment. Figure 4 shows that node 30 converges to -1.25 dBs when running the continuous algorithm and -3 dBs in the discretised case.

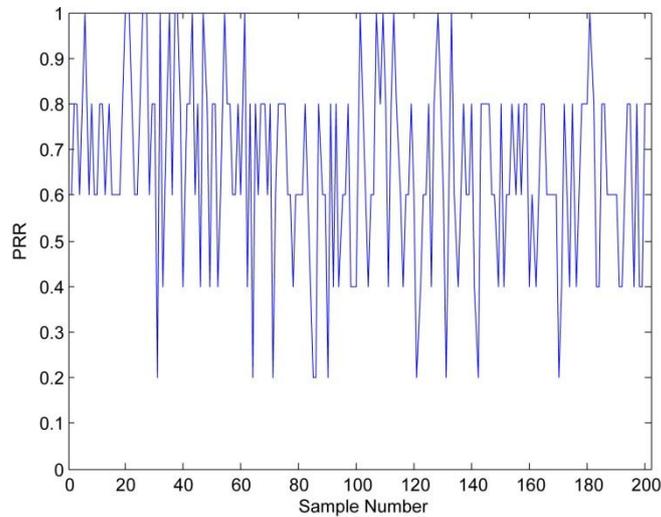

Figure 5. PRR of Continuous Tx

To further elaborate on the high Tx values we see in figures 5,6,7 that the PRR values of all three cases are very high In the continuous case PRR values are located between 40 % and 80% with the exception of certain bursts that go to 100 % and 20 %.

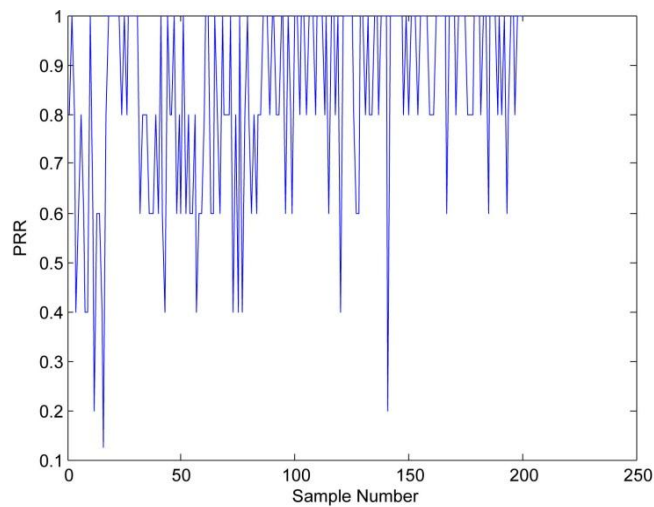

Figure 6. PRR of Discretised Tx

When Tx is discretised to a lower value as we saw previously, PRR values seem to go higher, 60 - 100 % except from a minor set of values that go quite low (close to 10 %). Of course, full power exhibits the highest PRR values. However, given the simulation environment we needed to verify our findings in a testbed environment. Thus, we decided to put the discretised version of our algorithm on Indriya.

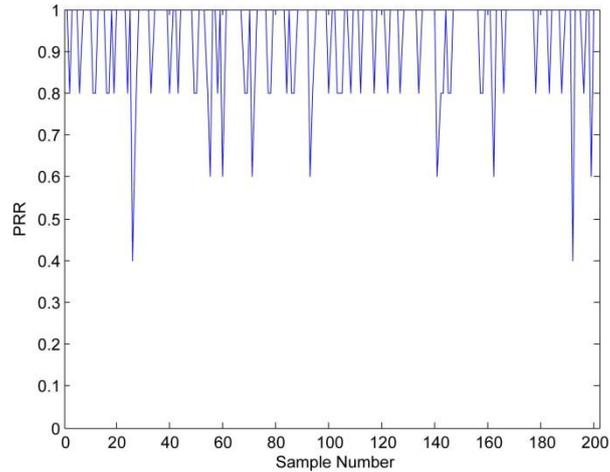

Figure 7. PRR of Full Tx

## 7. TESTBED RESULTS

In order to further investigate our simulation results, we decided to run the game-theoretic power control and the full transmission power algorithms with CTP on all available nodes at Indriya. Our configuration allowed 3 re-transmissions by CTP and a message has been send every 2 seconds to another node. We run both algorithms for a period of 1 hour in order to test their behaviour. Initially, our observations showed that the discretised power control algorithm exhibits 86.8% average PRR, while the algorithm running with full Tx 96.1% as can be seen in figure 8. In order to reflect on the effect of the transmission power change on our links, we constructed the Empirical Cumulative Function (CDF) which can be seen in figure 9. The CDF showed that the full Tx algorithm showed better intermediate links; however, our game theoretic power control algorithm does not perform very badly, it has a better probability of forming good quality links. Moreover, the full Tx algorithm has 88% of the links exhibiting PRR over 80% and approximately 12% intermediate PRR (value between 30-80%). On the other hand, our game theoretic power control algorithm comprises 75% of the links with good, 18% intermediate and 7% bad qualities.

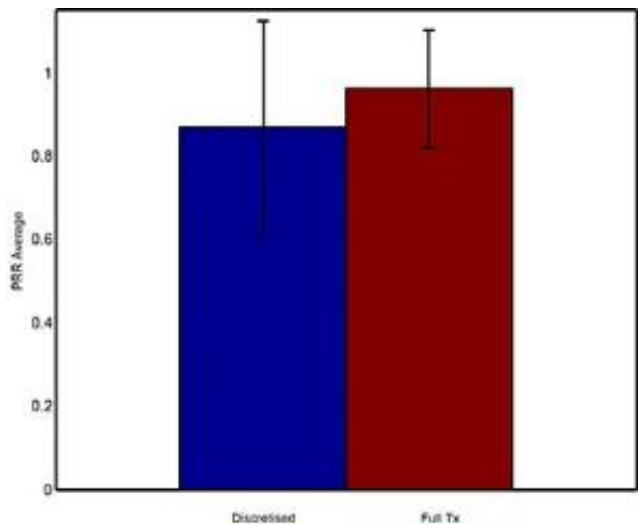

Figure 8. PRR average of discretised and Full Tx

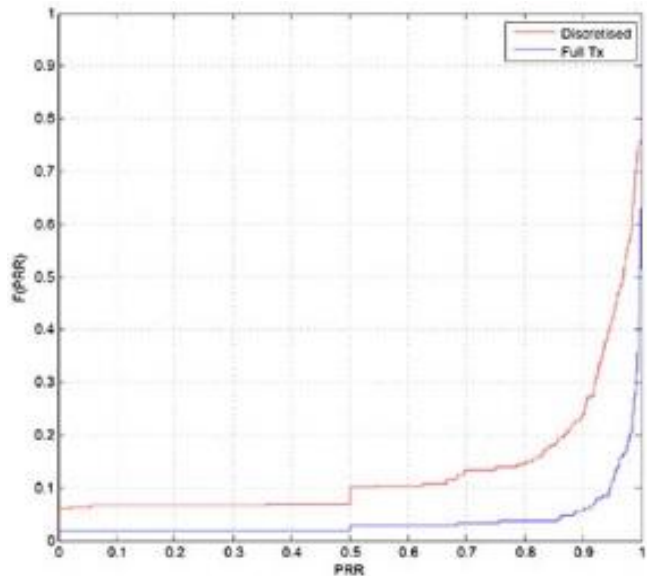

Figure 9. CDF of discretised and Full Tx Link Quality

Furthermore, we derived average relative Tx values for both algorithms in the same way as with our simulation results. This gave us an indication of the percentage of energy efficiency of our power control algorithm. The results showed that our algorithm saved approximately 11% energy. Figure 10 shows this result, which in conjunction with the quite high PRR that our approach accomplishes gives us an indication of the efficiency of our approach. According to the CDF we see that nodes exhibit very high PRRs and we can derive that they can do so with not the highest level of transmission power levels.

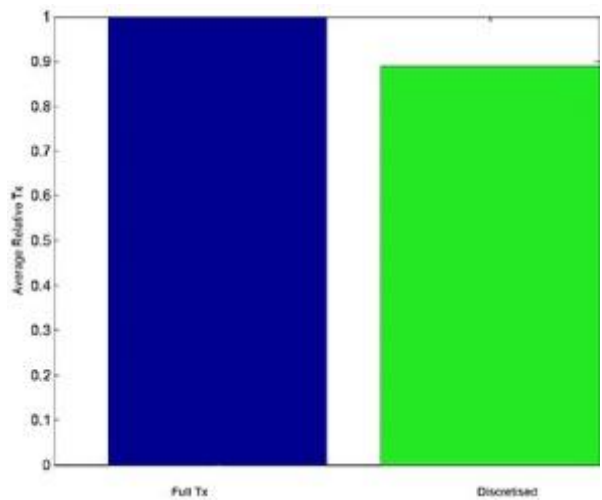

Figure 10. Average Relative Energy Consumption

Thereafter, we verified connectivity offline using method that determines whether the resulting graph is connected. Specifically, we created both algorithms' respective adjacency matrices. Thereafter, we used the matrices to find a zero eigenvalue. In the case that the corresponding eigenvector has 0s, then a sum of non-zero number of rows/columns of the adjacency is 0 [43]. Hence, the degrees of these nodes are 0 and the graph is disconnected. This is not the case with both the algorithms.

## 8. CONCLUSIONS

In this paper we addressed the problem of power control in wireless networks. We provided a model, which regards nodes of the network as selfish players trying to maximise their PRR and minimise energy consumption. We did so in a game theoretic manner. We designed a new metric that we named transmission reliability, which we used in our utility function. We proved that the game is a potential game, which means that it converges to a pure Nash equilibrium. We gave sufficient evidence that our game theoretic algorithm converges to the unique maximiser of the game.

We interfaced our power control algorithm with CTP, in order to check the difference in performance when using a state-of-the-art routing protocol. In order to achieve uniqueness, we played the game in a continuous fashion. Thereafter, we discretised the transmission power range to a granularity of 1 db, in order to approximate the unique solution. Our aim was to investigate the difference in PRR and Tx levels between the continuous and discretised versions of our algorithm, in order to later put in on a testbed.

We performed simulations in TOSSIM and we compared the two versions of our algorithm with nodes running CTP and transmitting at maximum transmission power. We saw that the difference in PRR was 3% between our two versions of the game theoretic algorithm. The difference in average relative energy consumption was less than 10% between our algorithm and CTP with full Tx.

This led us to investigate the aforementioned algorithms on a testbed. We experimented on Indriya and we saw a significant difference in average relative energy consumption between the discretised version of our algorithm and CTP with maximum Tx. The difference was approximately 11%. In terms of performance both algorithms exhibited PRR values higher than 80% and the difference between them was approximately 10%.

The future work includes interfacing our algorithm with other state-of-the-art routing protocols, such as the Backpressure Collection Protocol (BCP) [44], in order to determine its performance. We would like to investigate the temporal scheduling of BCP by providing spatial information with our power control.

**AUTHORS**

**Evangelos Spyrou** is a PhD candidate at the Aristotle University of Thessaloniki. He graduated from Northumbria University at Newcastle, where he obtained a BSc in Computing and MSc in Embedded Computer Systems Engineering. He has worked in various projects, including frigate software development for defence and natural disaster real applications. He is interested in Game Theory, stochastic optimisation and queues, as well as, in general, applying economic theories to networking problems.

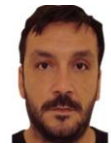

**Dimitris Mitrakos** is an Associate Professor at the Department of Electrical Engineering, School of Engineering, Aristotle University of Thessaloniki, Greece. His research interests include internet computing, multimedia communications, sensor and digital telemetry networks and distributed control and teleoperations systems. Dimitris has a Diploma in Electrical Engineering from Aristotle University of Thessaloniki, Aristotle University of Thessaloniki, an MSc in Communications Engineering from University of Manchester Institute of Science and Technology, a DIC in Signal Processing and a PhD in Electrical Engineering from University of London Imperial College of Science and Technology. In the recent past, he has been Vice-Chairman of the Electrical Engineering Department and Director of the Electronics and Computer Section of the Electrical Engineering Department of Aristotle University of Thessaloniki and Director of Postgraduate Studies of the Aristotle University of Thessaloniki Interdisciplinary Postgraduate Program in "Language and Communications Sciences". He acts regularly as member of Workprogramme Preparation Committees, Proposal Technical Evaluation Committees, Evaluation Process Assessment Committees and Project Technical Review Committees for a number of European Commission Programs.

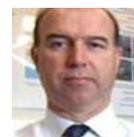